\documentclass[aps]{revtex4}

\begin{document}

\title{Existence of a Density Functional for an Intrinsic State}

\author{B. G. Giraud}
\affiliation{bertrand.giraud@cea.fr, Institut de Physique Th\'eorique, \\
DSM, CE Saclay, F-91191 Gif/Yvette, France}
\author{B. K. Jennings}
\affiliation{jennings@triumf.ca, TRIUMF, Vancouver BC, V6T2A3, Canada}
\author{B. R. Barrett}
\affiliation{ bbarrett@physics.arizona.edu, Department of Physics, \\ 
University of Arizona, Tucson, AZ 85721, USA}

\date{\today} 

\begin{abstract}

A generalization of the Hohenberg-Kohn theorem for finite systems proves 
the existence of a density functional (DF) for a symmetry violating intrinsic 
state, out of which a physical state with good quantum numbers can be 
projected.

\end{abstract}

\maketitle

\section{Introduction}

Density functional \cite{HK} theory (DFT) was initially defined for ground 
states. These have good quantum numbers. Every nuclear physicist knows that, 
for instance, the ground state of $^{20}$Ne is a $0^+$ and that its density 
is, thus, isotropic, not an ellipsoid. Every molecular physicist knows 
that, for instance, the ground state of the ammonia molecule is a good parity 
state, not just the pyramid described by the Born-Oppenheimer approximation. 
In particular, the nuclear DF must generate {\it spherical} solutions 
for the some thousand $0^+$ nuclear ground states, whether nuclei are 
intrinsically deformed or not. The same need for isotropic solutions extends 
to the non-local generalization of the DFT \cite{Gil} - \cite{DG}.
But the theory of rotational bands and/or parity vibrations, whether in 
nuclear, atomic or molecular physics, most often relates ground states to 
wave packets, often named ``intrinsic states'', which are symmetry breaking, 
namely do not transform in an irreducible representation (irrep) of the 
symmetry group $S$ of the Hamiltonian. Therefore, one may raise the question 
of DFT for intrinsic states rather than eigenstates. 

Given the physical Hamiltonian $H$ with its symmetry group $S,$ calculations 
providing a ``non S-irrep'' state as a solution for a minimum energy cannot 
be labelled as the result of ``the'' DF. Such a state, labelled intrinsic, 
is actually just a convenient wave packet, to be subsequently projected 
onto good quantum numbers to account for physical levels. Such intrinsic 
calculations should rather exhibit a special Hamiltonian, which might be 
called an intrinsic Hamiltonian, distinct from the physical one, if such 
calculations are to be legitimized. Or they should be interpreted 
as one variety of the Hartree-Fock, Hartree-Bogoliubov, etc. variational 
approaches. This is implicit or even explicit in calculations with 
an {\it energy density} functional, implying non-localities, see for 
instance \cite{JD} - \cite{GB}. Energy density and particle density are 
different concepts.

It turns out that the particle density which has been used for the foundation 
of DF theory mainly concerns eigenstates of the physical Hamiltonian, 
in principle at least, while the energy density, used for Skyrme force 
calculations in nuclear physics for instance, mainly provides intrinsic 
states. This note presents a {\it particle} DF theory for {\it intrinsic}
states, not for eigenstates of the Hamiltonian. We show how the physical 
Hamiltonian can be reconciled with the proper definition of a DF for an 
intrinsic state and how the resulting intrinsic state can be accepted as 
a useful wave packet, out of which states with good quantum numbers can be 
projected. In particular, in the case of molecules, our approach will consider 
both the electrons and the nuclei. Our intrinsic state can take into account
both kinds of degrees of freedom. Section II describes a functional out of 
which a variational principle derives for an intrinsic state, and out of 
which a DF for the intrinsic density is obtained. Section III gives an example 
of variational equations to be solved in practice. Section IV rewrites the 
formalism into a slightly simpler form. Section V contains a discussion of 
our result and suggests an ansatz for intrinsic Hamiltonians.

\section{Basic Formalism}

For a first argument, dealing with one kind of identical particles only, let 
$H$ be their physical Hamiltonian and $\phi,\ \langle \phi | \phi \rangle=1,$ 
be a trial wave packet, most often not transforming under an irrep of the 
symmetry group $S$ of $H.$ For instance, for fermions, $\phi$ may be an 
arbitrary Slater determinant, but we let $\phi$ be also a more general wave 
function, including some amount of correlations. States $\psi \propto P \phi$ 
with good quantum numbers can then be projected out of $\phi$ by a projector 
$P,$ a fixed operator. In the following, we shall systematically use the 
properties, $P^2=P$ and $[P,H]=0.$ 
It may happen that $\langle \phi | P | \phi \rangle$ vanishes, but such cases 
usually make a domain of zero measure in the usual variational domains, where 
$\phi$ evolves. In any case, since $H$ is an operator bounded from below,
the functional of $\phi,$ 
$\langle \phi | P H | \phi \rangle / \langle \phi | P | \phi \rangle,$ is 
bounded from below.
Embed now the system in an external, local field, $U=\sum_i u(r_i).$ The 
local, real potential $u$ is taken bounded from below, but is otherwise 
arbitrary. In particular, it may usually have none of the symmetries of $H.$
Then, given $u,$ the following functional of $\phi,$
\begin{equation}
{\cal F}[\phi] = \frac{\langle \phi | P H | \phi \rangle} 
{\langle \phi | P | \phi \rangle} + \langle \phi | U | \phi \rangle\, ,
\end{equation}
is bounded from below. To find the lowest energy with the quantum numbers 
specified by $P$ one can use a constrained search \cite{Levy}, in which one 
first considers only states that show a given density profile $\tau(r),$ then 
one lets $\tau$ vary,
\begin{equation}
{\rm Inf}_{\phi}\, \left[ 
\frac{\langle \phi | P H | \phi \rangle} {\langle \phi | P | \phi \rangle} +
\langle \phi | U | \phi \rangle \right]=
{\rm Inf}_{\tau}\, \left[\, \left({\rm Inf}_{\phi \rightarrow \tau}\, 
\frac{\langle \phi | P H | \phi \rangle} {\langle \phi | P | \phi \rangle}
\right) + \int dr\, \tau(r)\, u(r)\, \right].
\label{definF}
\end{equation}
The process goes in two steps, namely, i) a minimization within a given 
particle density profile, 
$\tau(r) \equiv \langle \phi | c^{\dagger}_r c_{r} | \phi \rangle,$ for $N$ 
particles, with $c^{\dagger}_r$ and $c_{r}$ the usual creation and 
annihilation operators at position $r,$ then, ii) a minimization with 
respect to the profile. The inner minimization clearly defines a DF, 
$F[\tau] \equiv {\rm Inf}_{\phi \rightarrow \tau}\, \left( 
\langle \phi | P H | \phi \rangle / \langle \phi | P | \phi \rangle \right).$ 

Actually, it is more general \cite{Lieb} - \cite{Leeu} to use many-body 
density matrices ${\cal B}$ in $N$-body space, meaning mixed as well as pure 
states, and yielding a density $\tau(r)$ in one-body space, 
\begin{equation}
{\rm Inf}_{\tau}\, \left[\, {\rm Inf}_{{\cal B} \rightarrow \tau}\, 
\left( \frac{ {\rm Tr}\, {\cal B}\, P H } { {\rm Tr}\, {\cal B}\, P } +
{\rm Tr}\, {\cal B}\, U \right)\, \right],
\end{equation}
but we shall use wave-functions in the following, namely 
${\cal B}=| \phi \rangle \langle \phi |,$ for obvious pedagogical 
reasons. We shall assume that this ${\rm Inf}_{\phi}$ actually defines an
absolute minimum, ${\rm Min}_{\phi},$ reached at some solution $\Phi$ of the 
corresponding variational principle. Moreover, we shall assume, temporarily 
at least, that the solution $\Phi$ is unique. Uniqueness is not obvious, 
however, if only because many $\phi$'s can give the same $P|\phi \rangle,$ 
and, when $u$ vanishes, this variational principle, Eq. (\ref{definF}), 
reduces to the well-known ``variation after projection'' \cite{DFGO} method 
for Hartree-Fock calculations for instance. 

Anyhow, $\tau$ and $u$ are clearly conjugate in a functional 
Legendre transform, with $\delta F / \delta \tau = -u.$
Finally, if $\rho(r)$ denotes the profile of $\Phi$ when 
$u \rightarrow 0,$ then the lowest energy with good quantum numbers is 
nothing but $F[\rho].$
The minimization, with respect to $\tau,$ of the functional, $F[\tau],$ 
provides simultaneously the density of the intrinsic (unprojected!) state 
and the projected energy. Notice, incidentally, that $F[\tau]$ depends on the 
choice of the variational set of trial functions $\phi$ where the ``inner 
minimization'' is performed. Furthermore, it obviously depends on $P.$

A more general argument is possible, with more than one kind of identical 
particles. Trial states can be, for instance, products of determinants, one 
for each kind of fermions, and permanents, one for each kind of bosons.  
Consider for instance the ammonia molecule, with i) its active
electrons, ii) its three protons and iii) its nitrogen ion. It is trivial 
to include a center-of-mass trap into $H$ to factorize into a spherical wave 
packet the center of mass motion of this self-bound system  and avoid 
translational degeneracy problems. The complete Hamiltonian $H$, trial states 
$\phi$ and density operators ${\cal B}$ depend on and describe simultaneously 
the electron, proton and nitrogen ion coordinates and momenta. A DF in just 
the electronic density space, however, results from the definition,
\begin{equation}
{\cal F}[\tau]={\rm Inf}_{{\cal B} \rightarrow \tau}\,
\frac{ {\rm Tr}\, {\cal B}\, P H } { {\rm Tr}\, {\cal B}\, P },
\end{equation}
where $P$ projects good quantum numbers for the whole system and traces are 
taken over all degrees of freedom, while $\tau$ is set as only an 
electronic density. Interactions between heavy degrees of freedom, between 
heavy and electronic ones, and between electrons, are taken into account by 
the trace in the numerator. No Born-Oppenheimer approximation is needed for 
this ``global'' definition. For the sake of simplicity, however, we return 
in the following to the case of one kind of particles only. Most 
considerations which follow have obvious generalization for multicomponent 
systems.

\section{Variational Equations}

Let $\delta \phi$ be an infinitesimal variation of the trial function in 
its allowed domain. Then, at first order, one obtains,
$$
\delta {\cal F} =
\frac{\langle \delta \phi | P H | \phi \rangle} 
{\langle \phi | P | \phi \rangle} +
\langle \delta \phi | U | \phi \rangle -
\frac{\langle \delta \phi | P | \phi \rangle\,
\langle \phi | P H | \phi \rangle} 
{( \langle \phi | P | \phi \rangle )^2}\ + 
$$
\begin{equation}
\ \ \  \ \ \ \  
\frac{\langle \phi | P H | \delta \phi \rangle} 
{\langle \phi | P | \phi \rangle} +
\langle \phi | U | \delta \phi \rangle -
\frac{\langle \phi | P | \delta \phi \rangle\,
\langle \phi | P H | \phi \rangle} 
{( \langle \phi | P | \phi \rangle )^2}\, .
\end{equation}
If one defines the ``gradient operator'',
\begin{equation}
{\cal G}=\frac{ P H } {\langle \phi | P | \phi \rangle} + U  -
\frac{ P \, \langle \phi | P H | \phi \rangle } 
{ ( \langle \phi | P | \phi \rangle )^2 }\, ,
\label{gradop} 
\end{equation}
then, obviously, 
$\delta {\cal F}=\langle \delta \phi | {\cal G} | \phi \rangle +
                 \langle \phi | {\cal G} | \delta \phi \rangle.$ Note, 
incidentally, that ${\cal G}$ is Hermitian.

At the minimum position $\Phi$, the variation $\delta {\cal F}$ vanishes 
for any $\delta \phi.$ Replace $\delta \phi$ by $i\, \delta \phi$ to see that 
the difference, $- \langle \delta \phi | {\cal G} | \Phi \rangle +
               \langle \Phi | {\cal G} | \delta \phi \rangle,$ vanishes
as well. Then, trivially, at $\Phi,$ both
$\langle \delta \phi | {\cal G} | \Phi \rangle$ and 
$\langle \Phi | {\cal G} | \delta \phi \rangle$ vanish simultaneously,
\begin{equation}
\langle \delta \phi | {\cal G} | \Phi \rangle=
\langle \Phi | {\cal G} | \delta \phi \rangle=0,\ \ \ \forall\, \delta \phi.
\label{statcond}
\end{equation}

In the special case of Slater determinants, let 
$| ph \rangle \equiv c^{\dagger}_p c_h | \Phi \rangle$ denote any 
particle-hole state built upon $| \Phi \rangle$ as the ``reference vacuum'' 
for quasi-particles. Here $c^{\dagger}$ and $c$ are the familiar fermionic 
creation and annihilation operators, respectively. Then the particle-hole 
matrix elements of ${\cal G}$ vanish,
\begin{equation}
\langle ph |\, {\cal G}\, | \Phi \rangle=0,\ \ \ \forall\, ph.
\label{phstatcond}
\end{equation}
As long as a solution of this stationarity condition, Eq. (\ref{phstatcond}),
is not reached, the matrix elements, $\langle ph | {\cal G} | \phi \rangle,$
define the direction of the gradient of ${\cal F}$ in the hyperplane tangent 
to the manifold of Slater determinants. A gradient descent algorithm,
$| \delta \phi \rangle= 
- \eta \sum_{ph} | ph \rangle\, \langle ph | {\cal G} | \phi \rangle,$
where $\eta$ is a small step parameter, then leads to the solution. Notice, 
however, that the $ph$ representation is covariant with $\phi.$ The $ph$ 
basis has to be recalculated at each step. Being state dependent, ${\cal G}$ 
must also be recalculated at each step.

\section{Similar theory, with a Lagrange multiplier}

The slightly complicated gradient operator, Eq. (\ref{gradop}), leads to a 
variational condition, Eq. (\ref{statcond}), which combines the matrix 
elements of three operators, namely $PH,$ $P$ and $U.$ Define the number
$\lambda=\langle \Phi | PH | \Phi \rangle/\langle \Phi | P | \Phi \rangle$
as a yet unknown Lagrange multiplier; it can be considered as an arbitrary 
parameter and shall be adjusted self-consistently later, when $\Phi$ is 
reached. Then Eq. (\ref{statcond}) also reads,
\begin{equation}
\langle \delta \phi |\, \left(\, PH -\lambda\, P + 
\langle \Phi | P | \Phi \rangle\, U\, \right)\, | \Phi \rangle =0.
\label{diago}
\end{equation}
If $\phi$ were completely unrestricted, namely if $\delta \phi$ were 
completely general, this equation, Eq. (\ref{diago}), would mean that $\Phi$ 
is an eigenstate of the operator ${\cal G}.$ Since intrinsic states are 
understood to belong to restricted sets of states, the result $\Phi$ is 
only an approximate eigenstate of ${\cal G}.$ 

To avoid the cumbersome coefficient, $\langle \Phi | P | \Phi \rangle,$ 
which multiplies $U,$ it is convenient to define the auxiliary operator,
\begin{equation}
{\cal H} = PH - \lambda\, P + W,
\label{auxil}
\end{equation}
where $W=\sum_i w(r_i)$ is, like $U,$ an arbitrary, local, real, external 
field, bounded from below. It is obvious that  ${\cal G}$ and ${\cal H}$ 
define a common solution $\Phi$ if $u$ and $w$ are suitably proportional to 
each other, $w=\langle \Phi | P | \Phi \rangle\, u.$ In the following, 
however, we set ${\cal H}$ {\it ab initio}. It is an operator bounded from 
below. We are interested in its ``almost ground state'' $\Xi$ and assume that 
this state is unique. A connection between a solution $\Xi$ in this section 
and a solution $\Phi$ in the previous section can easily be tested later.

Define again a constrained search for the lowest energy,
\begin{equation}
{\rm Inf}_{\phi}\, \langle \phi |\, {\cal H}\, | \phi \rangle = 
{\rm Inf}_{\tau} \left(\, {\rm Inf}_{\phi \rightarrow \tau}\, 
\langle \phi |\, {\cal H}\, | \phi \rangle\, \right) = {\rm Inf}_{\tau} \left(
F_{\lambda}[\tau] + \int dr\, \tau(r)\, w(r)\, \right),
\end{equation}
where the $\lambda$-dependent DF, $F_{\lambda},$ is defined as,
\begin{equation}
F_{\lambda}[\tau] \equiv {\rm Inf}_{\phi \rightarrow \tau}\, 
\langle \phi |\, (P H-\lambda P)\, | \phi \rangle.
\end{equation}
It is again convenient, for pedagogy at least, to assume that this 
${\rm Inf}_{\phi}$ induces an absolute minimum, reached at a position $\Xi$ in 
the variational space. The same assumption states that, given $\lambda,$ 
the absolute minimum of $F_{\lambda}[\tau]$ is 
\begin{equation}
F_{\lambda}[\sigma]=\langle \Xi | (P H - \lambda P) | \Xi \rangle,
\end{equation}
where $\sigma$ is the density of $\Xi.$ Let ${\cal E}$ denote this energy,
${\cal E}(\lambda) \equiv \langle \Xi | (P H - \lambda P) | \Xi \rangle.$
A simple manipulation then gives,
\begin{equation}
\frac{d {\cal E}}{d \lambda} = -\langle \Xi | P | \Xi \rangle.
\label{deri}
\end{equation}
A Legendre transform, using $\lambda$ and $\langle \Xi | P | \Xi \rangle$ as 
conjugate variables, is thus available to return the matrix element 
$\langle \Xi | P H | \Xi \rangle$ as a function of the matrix element 
$\langle \Xi | P | \Xi \rangle.$ Then one has just to locate the minimum of 
their ratio.

Note again that the theory depends on the variational space where $\phi$ 
evolves. But, in any case, one obtains simultaneously the density of 
$\overline \Xi,$ the best intrinsic state, and the energy of its projected 
state $P |\, \overline \Xi\, \rangle.$

\section{Summary, discussion and conclusion}

If only because of the need for spin densities \cite{GunLun} in the 
description of polarizable systems, the problem of symmetry conservation, or 
restoration, in DF theory has already received much attention 
in atomic and molecular physics \cite{Goer1} - \cite{Goer3}. It has been 
revisited here, in the spirit of the projected Hartree-Fock method with 
variation {\it after} projection \cite{DFGO}: a variational principle for 
the density of an intrinsic state, without symmetry, optimizes the energy 
of a state with good quantum numbers. The idea was already introduced in 
the context of particle number projection \cite{Ried}. We have shown in 
Secs. II and IV that our approach allows generalizations of the 
Hohenberg-Kohn existence theorem.

It can be stressed that the present approach is concerned with the density
of an intrinsic state, not that of an eigenstate. This is a major difference 
with all the other DF theories that we are aware of. Note, in particular, how 
our functional differs from a functional of a symmetrized \cite{Goer1} 
density . 

We showed in Sec. IV that a way to define the intrinsic Hamiltonian 
amounts to a linear combination, ${\cal H}=-\lambda\, P + P H,$ of the 
projector $P$ on the desired quantum numbers, and the laboratory Hamiltonian 
multiplied by that same $P.$ Here, a subtle question must be raised, 
that of the nature of the intrinsic state. The more flexible the trial 
functions for this state, the better the projected state and the lower the 
projected energy. However, full flexibility contradicts simplicity, and, 
moreover, uniqueness of the intrinsic state; many different packets 
$| \phi \rangle$ can give the same $P | \phi \rangle.$  Symmetry projection 
brings correlations which, therefore, {\it must} be absent from the 
intrinsic state. This is why variational domains for intrinsic states 
must {\it necessarily} be much narrower than the full Hilbert space. 

In practice, fortunately, intrinsic states are confined to non-linear, 
curved \cite{Courb} manifolds, such as coherent states, Slater determinants, 
etc., which do not make linear subspaces. The intrinsic state, therefore, 
is not an exact eigenstate of ${\cal H}.$ It just minimizes a related 
quantity, the projected energy. It must be concluded that DF 
theory for an intrinsic state necessarily depends on two factors, namely, 
i) obviously the quantum numbers to be projected out, but also ii) the
variational space retained for this intrinsic state.

\medskip

{\it Acknowledgements}: It is a pleasure for B.R.B. and B.G.G. to thank the 
TRIUMF Laboratory, Vancouver, B.C., Canada, for its hospitality, where part 
of this work was done. The Natural Science and Engineering Research Council 
of Canada is thanked for financial support. TRIUMF receives federal funding 
via a contribution agreement through the National Research Council of Canada.
B.R.B. also thanks Service de Physique Th\'eorique, Saclay, France, and the 
Gesellschaft f\"ur Schwerionenforschung mbh Darmstadt, Germany, for their 
hospitality, where parts of this work were carried out, and acknowledges 
partial support from the Alexander von Humboldt Stiftung and NSF Grant
PHY0555396. The contribution of two anonymous referees in making this paper 
clearer is also gratefully acknowledged.

\end{document}